\documentclass[12pt]{article}
\usepackage{amsmath}
\usepackage{amssymb}
\usepackage{amsfonts}
\usepackage{graphics}

\renewcommand{\theequation}{\arabic{section}.\arabic{equation}}
\newcommand{\field}[1]{\mathbb{#1}}
\newcommand{\N}{\field{N}}
\newcommand{\R}{\field{R}}
\newcommand{\Q}{{\cal Q}}

\title{Quasi-exactly solvable extensions of the Kepler-Coulomb potential on the sphere}
\author{C. Quesne\thanks{E-mail address: christiane.quesne@ulb.be}\\ 
{\small\sl Physique Nucl\'eaire Th\'eorique et Physique Math\'ematique,  Universit\'e Libre de Bruxelles,} \\ 
{\small\sl Campus de la Plaine CP229, Boulevard~du Triomphe, B-1050 Brussels, Belgium}}
\date{ }
\begin{document}
\baselineskip=22pt plus 1pt minus 1pt
%%%%%%%%%%%%%%%%%%%%%%%%%%%%%%%%%%%%%%%%%%%%%%%%%%%%%%%%%%
\maketitle
\begin{abstract}
We consider a family of extensions of the Kepler-Coulomb potential on a $d$-dimensional sphere and analyze it in a deformed supersymmetric framework, wherein the starting potential is known to exhibit a deformed shape invariance property. We show that the members of the extended family are also endowed with such a property, provided some constraint conditions relating the potential parameters are satisfied, in other words they are conditionally deformed shape invariant. Since, in the second step of the construction of a partner potential hierarchy, the constraint conditions change, we impose compatibility conditions between the two sets to build quasi-exactly solvable potentials with known ground and first-excited states. Some explicit results are obtained for the first three members of the family. We then use a generating function method, wherein the first two superpotentials, the first two partner potentials, and the first two eigenstates of the starting potential are built from some generating function $W_+(r)$ [and its accompanying function $W_-(r)$]. From the results obtained for the latter for the first three family members, we propose some formulas for $W_{\pm}(r)$ valid for the $m$th family member, depending on $m+1$ constants $a_0$, $a_1$, \ldots, $a_m$. Such constants satisfy a system of $m+1$ linear equations. Solving the latter allows us to extend the results up to the seventh family member and then to formulate a conjecture giving the general structure of the $a_i$ constants in terms of the parameters of the problem.
\end{abstract}


%
%========================================================================
%
\newpage

\section{Introduction}

The quantum oscillator problem in a $d$-dimensional space of constant curvature, which has arisen as a generalization of the classical nonlinear oscillator introduced by Mathews and Lakshmanan \cite{mathews}, has led to many studies \cite{carinena04a, carinena07a, carinena07b, carinena07c, carinena12, schulze, cq15}. Its exact solvability has been shown \cite{cq16} to be related to its deformed shape invariance (DSI) property in a deformed supersymmetric (DSUSY) framework \cite{bagchi}. Some rational extensions of the quantum oscillator on the sphere have also been proved to be exactly solvable \cite{cq16} with bound-state wavefunctions expressible in terms of exceptional orthogonal polynomials (see, e.g., Ref.~\cite{gomez} and references quoted therein), instead of classical orthogonal polynomials for the oscillator alone.\par
%
%-----------------------------------------------------------------------------------------------------------------
%
Other types of extensions of the oscillator in a $d$-dimensional constant-curvature space have been constructed and shown to be quasi-exactly solvable (QES), i.e., with only a finite number of eigenstates that can be found by algebraic means, while the remaining ones remain unknown. Whereas the simplest QES problems are characterized by a hidden sl(2, $\R$) algebraic structure (see, e.g., Ref.~\cite{turbiner} and references quoted therein) and are connected with polynomial solutions of the Heun equation \cite{ronveaux}, those considered for the oscillator in curved space are more complicated QES problems related to generalizations of the Heun equation. In a first approach, use has been made \cite{cq17} of the functional Bethe ansatz method \cite{zhang}. In a second approach \cite{cq18a}, infinite families of QES extensions of the oscillator in curved space with known ground and first-excited states have been constructed by combining a conditionally deformed shape invariance (CDSI) method, generalizing a procedure known for problems in an Euclidean space \cite{chakrabarti, bera}, with a generating function method developed both in SUSY \cite{tkachuk} and DSUSY \cite{voznyak} backgrounds.\par
%
%--------------------------------------------------------------------------------------------------------------
% 
In addition to the oscillator, another potential considered since the beginning of quantum mechanics has been the Kepler-Coulomb (KC) one. Its study on the sphere has indeed a long history, since it dates back to Schr\"odinger \cite{schrodinger}, Infeld \cite{infeld}, and Stevenson \cite{stevenson}. Kalnins et al.\ discussed the separation of variables of the Schr\"odinger equation with a KC potential on the sphere \cite{kalnins}. Higgs \cite{higgs} and Leemon \cite{leemon} analyzed the characteristics of the potential on the $d$-dimensional sphere. Since then, a lot of works have been devoted to the subject from several viewpoints, such as superintegrability or addition of some nonlinear terms (see, e.g., \cite{carinena05, ballesteros, ranada, hakobyan, carinena21a, carinena21b} and references quoted therein).\par
%
%-------------------------------------------------------------------------------------------------------
%
In particular, the quantum KC problem in a constant-curvature space was shown to exhibit a DSI property in DSUSY and some exactly-solvable rational extensions of the KC potential in a hyperbolic space were constructed with bound-state wavefunctions expressed in terms of exceptional orthogonal polynomials \cite{cq16}.\par
%
%-------------------------------------------------------------------------------------------------
%
The purpose of the present paper is to apply the approaches developed in Ref.~\cite{cq18a} to build extensions of the KC potential on the $d$-dimensional sphere with known ground and first-excited states. As the problem turns out to be more complicated than the previous one for the oscillator, we plan to exploit successively the use of the CDSI method and that of the generating function one.\par
%
%---------------------------------------------------------------------------------------------------------------------
%
The plan of the paper is as follows. In Section~2, the Schr\"odinger equation for the KC potential on a $d$-dimensional sphere is reviewed, as well as its DSUSY description and the corresponding DSI property. In Section~3, the CDSI method is applied to the first three potentials of a family of extensions. In Section~4, the results obtained in Section~3 are re-interpreted in the generating function method language and a generalization to arbitrary members of the extension family is sketched. Finally, Section~5 contains the conclusion.\par
%
%===================================================================
%
\section{The Kepler-Coulomb problem on the sphere and deformed supersymmetry}

In Ref.~\cite{carinena04b}, it was shown that the Lagrangian of the classical nonlinear oscillator introduced by Mathews and Lakshmanan \cite{mathews} can be generalized to a $d$-dimensional Euclidean space, where it is defined by 
\begin{equation}
  L = \frac{1}{2}\frac{m}{1+\lambda r^2} \left(\sum_i v_i^2 + \lambda \sum_{i<j} (x_iv_j - x_jv_i)^2\right) - 
  {\cal V}(r), \qquad {\cal V}(r) =\frac{1}{2} m \alpha^2 \frac{r^2}{1+\lambda r^2}.  \label{eq:L}
\end{equation}
Here all summations run over $i,j = 1, 2, \ldots, d$, and $r^2 \equiv \sum_i x_i^2$, with $r$ running on $(0, + \infty)$ or $(0, 1/\sqrt{|\lambda|})$ according to whether the nonlinearity parameter $\lambda$ is positive or negative. It was also shown that to $L$ corresponds a Hamiltonian
\begin{equation}
  H = \frac{1}{2m}\left(\sum_i p_i^2 + \lambda \left(\sum_i x_i p_i\right)^2\right) + {\cal V}(r) = 
  \frac{1}{2m}\left((1+\lambda r^2) \sum_i p_i^2
  - \lambda \sum_{i<j} J_{ij}^2\right) + {\cal V}(r),  \label{eq:NL-class}
\end{equation}
where $J_{ij}=x_i p_j - x_j p_i$. Such a Hamiltonian in Euclidean space was then proved to be interpretable as describing the Hamiltonian of an oscillator in a constant-curvature space with the curvature $\kappa$ related to the nonlinearity parameter $\lambda$ through the relation $\kappa = - \lambda$.\par
%
%---------------------------------------------------------------------------------------------------------
%
A similarr approach was later on applied \cite{carinena05} to the case where the nonlinear oscillator considered in (\ref{eq:L}) and (\ref{eq:NL-class}) is replaced by a nonlinear KC potential  
\begin{equation}
  {\cal V}(r) = - \frac{Q}{r}\sqrt{1+\lambda r^2}, \qquad Q>0.  \label{eq:KC-pot}
\end{equation}
The latter is also amenable to an interpretation  as a KC potential in a space of constant curvature $\kappa = - \lambda$, generalizing to $d$ dimensions the one considered in three dimensions by Schr\"odinger \cite{schrodinger}, Infeld \cite{infeld}, and Stevension \cite{stevenson}.\par
%
%-------------------------------------------------------------------------------------------------------------
%
To quantize (\ref{eq:NL-class}) with ${\cal V}(r)$ given in either (\ref{eq:L}) or (\ref{eq:KC-pot}), one replaces $\sqrt{1+\lambda r^2} \, p_i$ and $J_{ij}$ by the operators $- {\rm i}\sqrt{1+ \lambda r^2} \partial/\partial x_i$ and $\hat{J}_{ij} = - {\rm i}(x_i \partial/\partial x_j - x_j \partial/\partial x_i)$, respectively, thus leading to
\begin{equation}
\begin{split}
  {\hat H} &= - \left((1+\lambda r^2)\hat{\Delta} + \lambda r \frac{\partial}{\partial r} + \lambda {\hat J}^2
       \right) + {\cal V}(r) \\
  &= - \left((1+\lambda r^2)\frac{\partial^2}{\partial r^2} + (d-1+d\lambda r^2)\frac{\partial}{\partial r} 
       - \frac{{\hat J}^2}{r^2}\right) + {\cal V}(r),
\end{split}. \label{eq:NL-quant}
\end{equation}
where ${\hat J}^2 \equiv \sum_{i<j} \hat{J}_{ij}^2$ and $\hat{\Delta}$ denotes the Laplacian in a $d$-dimensional Euclidean space (note that here we take units wherein $\hbar = 2m =1$). In the present paper, we are going to consider more specifically the KC potential (\ref{eq:KC-pot}) on a sphere, which means that $\kappa = - \lambda>0$, leaving the other case $\kappa = - \lambda < 0$ for a further study.\par
%
%--------------------------------------------------------------------------------------------------
%
The Schr\"odinger equation corresponding to (\ref{eq:NL-quant}) is separable in hyperspherical coordinates and gives rise to the radial equation
\begin{equation}
  \left(-(1-\kappa r^2) \frac{d^2}{dr^2} - (d-1-d\kappa r^2)\frac{1}{r} \frac{d}{dr} + \frac{l(l+d-2)}{r^2}
  - \frac{Q}{r} \sqrt{1-\kappa r^2} - {\cal E}\right) R(r) = 0,  \label{eq:SE}
\end{equation}
where $l=0$, 1, 2, \ldots\ is the angular momentum quantum number, $Q>0$, and the radial variable $r$ runs over $(0,1/\sqrt{\kappa})$. The differential operator in (\ref{eq:SE}) being formally self-adjoint with respect to $(1-\kappa r^2)^{-1/2} r^{d-1}dr$, bound-state radial wavefunctions $R(r)$ should therefore be normalizable with respect to $(1-\kappa r^2)^{-1/2} r^{d-1}dr$.\par
%
%----------------------------------------------------------------------------------------------------------------
%
Equation~(\ref{eq:SE}) can be rewritten in an alternative form as
\begin{equation}
  \left(- \sqrt{f(r)} \frac{d}{dr} f(r) \frac{d}{dr} \sqrt{f(r)} + V(r) - E\right) \psi(r)=0,  \label{eq:SE-def}
\end{equation}
where
\begin{equation}
\begin{split}
  & f(r) = \sqrt{1-\kappa r^2}, \\
  & V(r) = V(r;l,Q) = \frac{L(L+1)}{r^2} - \frac{Q}{r}f(r), \quad L = l + \frac{d-3}{2}, \\
  & E = {\cal E} + \frac{1}{4} \kappa (d-1)^2, \\
  & \psi(r) = r^{(d-1)/2} f^{-1/2}(r) R(r),
\end{split}. \label{eq:SE-def-1}
\end{equation}
and can then be interpreted as a deformed Schr\"odinger equation, written in terms of a deformed radial momentum $\hat{\pi}_r = \sqrt{f(r)} (- {\rm i} d/dr) \sqrt{f(r)}$, or as a position-dependent mass (PDM) Schr\"odinger equation
\begin{equation}
  \left(- m^{-1/4}(r) \frac{d}{dr} m^{-1/2}(r) \frac{d}{dr} m^{-1/4}(r) + V(r) - E\right) \psi(r) = 0
\end{equation}
with $m(r) = 1/f^2(r)$ \cite{cq04}, the ordering of the PDM $m(r)$ and the differential operator $d/dr$ being that of Mustafa and Mazharimousavi \cite{mustafa}. Bound-state wavefunctions correspond to functions $\psi(r)$ normalizable with respect to $dr$ on the interval $(0, 1/\sqrt{\kappa})$.\par
%
%-----------------------------------------------------------------------------------------------------------------
%
It turns out \cite{cq16} that with $f(r)$ and $V(r)$ given in (\ref{eq:SE-def-1}), Eq.~(\ref{eq:SE-def}) has an infinite number of bound-state eigenvalues, given by
\begin{equation}
  E_{n_r}(l,Q) = - \frac{Q^2}{4n^2} + \kappa n^2, \qquad n = n_r + l + \frac{d-1}{2}, \qquad n_r = 0, 1, 2, \ldots,
\end{equation}
with corresponding wavefunctions\footnote{Slightly different, but equivalent, forms were used in Ref.~\cite{cq16}.} expressed in terms of Jacobi polynomials with complex indices and argument,
\begin{equation}
  \psi_{n_r}(r; l,Q) \propto r^n f^{-1/2} \exp\left(- \frac{Q}{2n\sqrt{\kappa}} \arcsin(\sqrt{\kappa}r)\right)
  P_{n_r}^{\left(-n + \frac{{\rm i}Q}{2n\sqrt{\kappa}}, - n -\frac{{\rm i}Q}{2n\sqrt{\kappa}}\right)}
      \left(\frac{{\rm i}f}{\sqrt{\kappa}r}\right),
\end{equation}
or in terms of Romanovski polynomials with real indices and argument \cite{romanovski, raposo},\footnote{Romanovski polynomials are also termed pseudo-Jacobi polynomials \cite{koekoek}.}
\begin{equation}
  \psi_{n_r}(r; l,Q) \propto r^n f^{-1/2} \exp\left(- \frac{Q}{2n\sqrt{\kappa}} \arcsin(\sqrt{\kappa}r)\right)
  R_{n_r}^{\left(\frac{Q}{n\sqrt{\kappa}}, - n + 1\right)}\left(\frac{f}{\sqrt{\kappa}r}\right).
\end{equation}
\par
%
%-------------------------------------------------------------------------------------------------------------
%
A deformed Schr\"odinger equation, such as (\ref{eq:SE-def}), can be discussed in terms of DSUSY \cite{bagchi}. In the case of unbroken DSUSY, after introducing a rescaled potential
\begin{equation}
  V_1(r) = V(r) - E_0,  \label{eq:rescaled-V}
\end{equation}
defined in terms of the ground-state energy $E_0$ of Eq.~(\ref{eq:SE-def}), one deals with a pair of partner Hamiltonians, defined by
\begin{equation}
  \hat{H}_{1,2} = \hat{\pi}_r^2 + V_{1,2}(r) + E_0, \qquad V_{1,2}(r) = W^2(r) \mp f(r) \frac{dW}{dr},
  \label{eq:partner-H}
\end{equation}
in terms of a superpotential
\begin{equation}
  W(r) = - f \frac{d}{dr} \log \psi_0(r) - \frac{1}{2} \frac{df}{dr}.
\end{equation}
Here $\psi_0(r)$ denotes the ground-state wavefunction of (\ref{eq:SE-def}), expressed in terms of the superpotential as
\begin{equation}
  \psi_0(r) \propto f^{-1/2} \exp\left(- \int^r \frac{W(r')}{f(r')} dr'\right).  \label{eq:gs}
\end{equation}
\par
%
%---------------------------------------------------------------------------------
%
The two partner Hamiltonians can be written as
\begin{equation}
  \hat{H}_1 = \hat{A}^+ \hat{A}^- + E_0, \qquad \hat{H}_2 = \hat{A}^- \hat{A}^+ + E_0,
\end{equation}
in terms of two first-order differential opearators
\begin{equation}
  \hat{A}^{\pm} = \mp \sqrt{f(r)} \frac{d}{dr} \sqrt{f(r)} + W(r).  \label{eq:A}
\end{equation}
They intertwine with $\hat{A}^+$ and $\hat{A}^-$ as
\begin{equation}
  \hat{A}^- \hat{H}_1 = \hat{H}_2 \hat{A}^-, \qquad \hat{A}^+ \hat{H}_2 = \hat{H}_1 \hat{A}^+.
\end{equation}
The operator $\hat{A}^-$ annihilates the ground-state wavefunction (\ref{eq:gs}) of $\hat{H}_1$, whereas $\hat{A}^+$ transforms the ground-state wavefunction $\psi'_0(r)$ of $\hat{H}_2$ into the first-excited state wavefunction $\psi_1(x)$ of $\hat{H}_1$.\par
%
%------------------------------------------------------------------------------------------------------------------
%
On considering $\hat{H}_2$ as a new starting Hamiltonian, one may in principle iterate the procedure and obtain a DSUSY pair of partner Hamiltonians
\begin{equation}
  \hat{H}'_{1,2} = \hat{\pi}_r^2 + V'_{1,2}(r) + E'_0, \qquad V'_{1,2}(r) = W^{\prime2}(r) \mp f(r)
  \frac{dW'}{dr},  \label{eq:partner-H-bis}
\end{equation}
where
\begin{equation}
  V'_1(r) + E'_0 = V_2(r) + E_0.  \label{eq:V'-V}
\end{equation}
The first-excited state wavefunction $\psi_1(r)$ of $\hat{H}_1$, with energy $E_1 = E'_0$, can then be obtained from the ground-state wavefunction of $\hat{H}'_1 = \hat{H}_2$, given by
\begin{equation}
  \psi'_0(r) \propto f^{-1/2} \exp\left(-\int^r \frac{W'(r')}{f(r')} dr'\right),  \label{eq:gs-bis}
\end{equation}
through the equation
\begin{equation}
  \psi_1(r) \propto \hat{A}^+ \psi'_0(r).  \label{eq:es}
\end{equation}
\par
%
%-------------------------------------------------------------------------------------------------------------
%
In the case of the KC potential on the sphere of Eq.~(\ref{eq:SE-def-1}), the superpotential is given by \cite{cq16}
\begin{equation}
  W(r) = W(r;l,Q) = - \frac{L+1}{r} f(r) + \frac{Q}{2L+2}
\end{equation}
and the two partner potentials are
\begin{equation}
  V_1(r;l,Q) = \frac{L(L+1)}{r^2} - \frac{Q}{r} f(r) - E_0
\end{equation}
and
\begin{equation}
  V_2(r;l,Q) = \frac{(L+1)(L+2)}{r^2} - \frac{Q}{r} f(r) - E_0 = V_1(r;l+1,Q).
\end{equation}
Hence, the partner is similar in shape and its parameters are obtained by translation, i.e., $l \to l+1$, $Q\to Q$. This means that the KC potential on the sphere is DSI, so that a whole hierarchy of Hamiltonians can be straightforwardly constructed, and this explains the exact solvability of the starting Schr\"odinger equation.\par
%
%---------------------------------------------------------------------------------------------------------------------
%
In the following Sections, we plan to consider some extensions of the potential $V(r)$ of Eq.~(\ref{eq:SE-def-1}), namely
\begin{equation}
  V(r) = \frac{L(L+1)}{r^2} - \frac{Q}{r}f + \kappa \sum_{k=1}^m \left(B_{2k-1}\frac{r}{f^{2k-1}} + B_{2k}
  \frac{1}{f^{2k}}\right),  \label{eq:ext-V}
\end{equation}
with $L = l + \frac{d-3}{2}$, $Q>0$, $f(r) = \sqrt{1-\kappa r^2}$, and $B_1$, $B_2$, \ldots, $B_{2m}$ are $2m$ real parameters, such that $B_{2m}>0$. The additional terms then break the DSI property of the KC potential, so that the resulting Schr\"odinger equation may at most be QES.\par
%
%================================================================
%
\section{Conditionally deformed shape invariance symmetry method}

\setcounter{equation}{0}

In the present Section, we plan to consider successivaly the first few members of the potential family (\ref{eq:ext-V}) and apply the CDSI method of Ref.~\cite{cq18a}.\par
%
%+++++++++++++++++++++++++++++++++++++++++++++++++++++++++++++++++
%
\subsection{First extension}

{}For $m=1$, the potential (\ref{eq:ext-V}) reduces to 
\begin{equation}
  V(r) = \frac{L(L+1)}{r^2} - \frac{Q}{r} f + \kappa B_1 \frac{r}{f} + \kappa B_2 \frac{1}{f^2}, \qquad B_2>0,
  \label{eq:ext-V-1}
\end{equation}
and depends on the four parameters $L$, $Q$, $B_1$, and $B_2$. Let us assume a superpotential of the type
\begin{equation}
  W(r) = \frac{\xi}{r} f + \eta + \zeta \frac{r}{f}, \qquad \xi\le 0, \qquad \zeta>0,
\end{equation}
depending on three parameters $\xi$, $\eta$, and $\zeta$. As shown in (\ref{eq:partner-H}), in DSUSY the rescaled potential (\ref{eq:rescaled-V}) is represented by $V_1(r) = W^2 - f dW/dr$. With the extended potential (\ref{eq:ext-V-1}), this Riccati equation gives rise to the system of equations
\begin{equation}
\begin{split}
  &\xi(\xi+1) = L(L+1), \\
  &- \kappa \xi^2 + 2\xi\zeta + \eta^2 - \frac{\zeta^2}{\kappa} = - E_0, \\
  &2\xi\eta = -Q, \\
  &2\eta\zeta = \kappa B_1, \\
  &\frac{\zeta}{\kappa} \left(\frac{\zeta}{\kappa}-1\right) = B_2
\end{split}. \label{eq:Riccati-1}
\end{equation}
for the three unknowns $\xi$, $\eta$, $\zeta$. The first, third, and fourth of these equations lead to the values of the latter
\begin{equation}
  \xi = -L-1, \qquad \eta = \frac{Q}{2(L+1)}, \qquad \zeta = \frac{\kappa(L+1)B_1}{Q}.  \label{eq:W-para}
\end{equation}
Since $\zeta>0$, we must have $B_1>0$. From the second equation of (\ref{eq:Riccati-1}), the ground-state energy $E_0$ is obtained in the form
\begin{equation}
  E_0 = \kappa(L+1)^2 - \frac{Q^2}{4(L+1)^2} + \frac{\kappa(L+1)B_1}{Q} \left(\frac{(L+1)B_1}{Q}
  +2L+2\right).
\end{equation}
The last equation of (\ref{eq:Riccati-1}) then provides a constraint
\begin{equation}
  B_2 = \frac{(L+1)B_1}{Q} \left(\frac{(L+1)B_1}{Q} - 1\right),  \label{eq:constraint-1}
\end{equation}
connecting the potential parameters. The condition $B_2>0$ imposes that $B_1>Q/(L+1)$. To $E_0$ corresponds the wavefunction (\ref{eq:gs}), which can be written as
\begin{equation}
  \psi_0(r) \propto r^{-\xi} \exp\left(- \frac{\eta}{\sqrt{\kappa}} \arcsin(\sqrt{\kappa}r)\right) 
  f^{\frac{\zeta}{\kappa} - \frac{1}{2}},
\end{equation}
with $\xi$, $\eta$, and $\zeta$ given in (\ref{eq:W-para}).\par
%
%------------------------------------------------------------------------------------------------------------------
%
The partner $V_2(r) = W^2 + fdW/dr$ of $V_1(r)$ in DSUSY can be written as
\begin{equation}
  V_2(r) = \frac{L'(L'+1)}{r^2} - \frac{Q'}{r}f + \kappa B'_1 \frac{r}{f} + \kappa B'_2 \frac{1}{f^2} + R,
\end{equation}
in terms of some new parameters $L'$, $Q'$, $B'_1$, $B'_2$, and a constant $R$, which must satisfy the system of equations
\begin{equation}
\begin{split}
  &L'(L'+1) = \xi(\xi-1), \\
  &R = - \kappa \xi^2 + 2\xi \zeta + \eta^2 - \frac{\zeta^2}{\kappa}, \\
  &-Q' = 2\xi\eta, \\
  &\kappa B'_1 = 2\eta\zeta, \\
  &B'_2 = \frac{\zeta}{\kappa} \left(\frac{\zeta}{\kappa}+1\right).
\end{split}
\end{equation}
On using some results obtained above, we get
\begin{equation}
  L' = L+1, \qquad Q' = Q, \qquad B'_1 = B_1, \qquad B'_2 = B_2 + 2 \frac{(L+1)B_1}{Q},
\end{equation}
as well as
\begin{equation}
  R = - E_0.
\end{equation}
This shows that the starting potential $V_1(r)$ is DSI, but such a deformed shape invariance is not unconditionally valid since the constraint (\ref{eq:constraint-1}) must be satisfied. The potential is therefore only CDSI.\par
%
%-------------------------------------------------------------------------------------------------------
%
We may now try to repeat the procedure by taking the partner $V_2(r)$ as a starting potential $V'(r) = V_2(r)$ with ground-state energy $E'_0$. We therefore consider a rescaled potential $V'_1(r) = V'(r) + E_0 - E'_0$ and a new superpotential
\begin{equation}
  W'(r) = \frac{\xi'}{r}f + \eta' + \zeta'\frac{r}{f}, \qquad \xi' \le 0, \qquad \zeta'>0.
\end{equation}
By proceeding as in the first step, we obtain for the new parameters
\begin{equation}
  \xi' = -L-2, \qquad \eta' = \frac{Q}{2(L+2)}, \qquad \zeta' = \frac{\kappa(L+2)B_1}{Q},
\end{equation}
and for the new ground-state energy
\begin{equation}
  E'_0 = \kappa(L+2)^2 - \frac{Q^2}{4(L+2)^2} + \frac{\kappa(L+2)B_1}{Q} \left(\frac{(L+2)B_1}{Q} + 2L +4
  \right).
\end{equation}
We also get a new constraint
\begin{equation}
  B_2 = \frac{(L+2)B_1}{Q} \left(\frac{(L+2)B_1}{Q} - 1\right) - 2\frac{(L+1)B_1}{Q},  \label{eq:constraint-2}
\end{equation}
relating the potential parameters. The ground-state wavefunction of the partner reads
\begin{equation}
  \psi'_0(r) \propto r^{-\xi'} \exp\left(- \frac{\eta'}{\sqrt{\kappa}} \arcsin(\sqrt{\kappa}r)\right)
  f^{\frac{\zeta'}{\kappa} - \frac{1}{2}}.
\end{equation}
\par
%
%---------------------------------------------------------------------------------------------------
%
The two constraints (\ref{eq:constraint-1}) and (\ref{eq:constraint-2}) are compatible provided
\begin{equation}
  B_1 = Q.
\end{equation}
Then
\begin{equation}
  B_2 = L(L+1), \qquad \zeta = \kappa(L+1),
\end{equation}
and the potential
\begin{equation}
  V(r) = \frac{L(L+1)}{r^2} - \frac{Q}{r}f + \kappa Q \frac{r}{f} + \kappa \frac{L(L+1)}{f^2},
  \label{eq:ext-V-1-bis}
\end{equation}
with corresponding superpotentials
\begin{equation}
\begin{split}
  &W(r) = - \frac{L+1}{r}f + \frac{Q}{2(L+1)} + \kappa(L+1) \frac{r}{f}, \\
  &W'(r) = - \frac{L+2}{r}f + \frac{Q}{2(L+2)} + \kappa(L+2) \frac{r}{f},
\end{split}. \label{eq:ext-W}
\end{equation}
has a ground state and a first-excited state whose energies are given by
\begin{equation}
  E_0 = 4\kappa(L+1)^2 - \frac{Q^2}{4(L+1)^2}, \qquad E_1 = E'_0 = 4\kappa(L+2)^2 - \frac{Q^2}{4(L+2)^2}.
 \label{eq:E}
\end{equation}
\par
%
%------------------------------------------------------------------------------------------------------
%
The ground-state wavefunction of potential (\ref{eq:ext-V-1-bis}) is given by
\begin{equation}
  \psi_0(r)  \propto r^{L+1} \exp\left(- \frac{Q}{2\sqrt{\kappa}(L+1)} \arcsin(\sqrt{\kappa}r)\right)
  f^{L+\frac{1}{2}},
\end{equation}
which is a normalizable function on $(0,1/\sqrt{\kappa})$. Moreover, its first excited-state wavefunction $\psi_1(r)$ can be obtained from the partner ground-state wavefunction
\begin{equation}
  \psi'_0(r)  \propto r^{L+2} \exp\left(- \frac{Q}{2\sqrt{\kappa}(L+2)} \arcsin(\sqrt{\kappa}r)\right)
  f^{L+\frac{3}{2}}
\end{equation}
by acting with the operator $\hat{A}^+$, defined in (\ref{eq:A}), which can be rewritten as  
\begin{equation}
  \hat{A}^+ = - f \frac{d}{dr} - \frac{L+1}{r} f + \frac{Q}{2(L+1)} + \kappa \left(L+\frac{3}{2}\right) \frac{r}{f}.
\end{equation}
The result reads
\begin{align}
  \psi_1(r) &\propto r^{L+1} \exp\left(- \frac{Q}{2\sqrt{\kappa}(L+2)} \arcsin(\sqrt{\kappa} r)\right) 
  f^{L+\frac{1}{2}} \nonumber\\
  &\quad \times \left(-1 + 2\kappa r^2 + \frac{Q}{2(L+1)(L+2)} rf\right),
\end{align}
which is also a normalizable function on $(0, 1/\sqrt{\kappa})$, with a single zero
\begin{equation}
  r_0 = \frac{1}{\sqrt{2\kappa}} \left(1 - \frac{Q}{\sqrt{Q^2 + 16\kappa (L+1)^2 (L+2)^2}}\right)^{1/2}
\end{equation}
on this interval, as it should be.\par
%
%---------------------------------------------------------------------------------------------
%
Note that, in the following, it will prove convenient to write the results in terms of the parameters $L$ and $\Q = Q/[2(L+1)(L+2)]$, instead of $L$ and $Q$. With such conventions, we get
\begin{equation}
\begin{split}
  & B_1 = 2(L+1)(L+2) \Q, \qquad B_2 = L(L+1), \\
  & W(r) = - \frac{L+1}{r} f + (L+2) \Q + \kappa (L+1) \frac{r}{f}, \\
  & W'(r) = - \frac{L+2}{r} f + (L+1)\Q + \kappa (L+2) \frac{r}{f}, \\
  & E_0 = 4\kappa (L+1)^2 - (L+2)^2 \Q^2, \qquad E_1 = 4\kappa (L+2)^2 - (L+1)^2 \Q^2.
\end{split}
\end{equation}
\par
%
%--------------------------------------------------------------------------------------------------------------
% 
In Fig.~1, an example of extended potential (\ref{eq:ext-V-1-bis}) is plotted and compared with the starting KC potential. The corresponding (unnormalized) $\psi_0(r)$ and $\psi_1(r)$ of the former are displayed in Fig.~2.\par
%
%-----------------------------------------------------------------------------------------------------------
%
\begin{figure}[h]
\begin{center}
\includegraphics{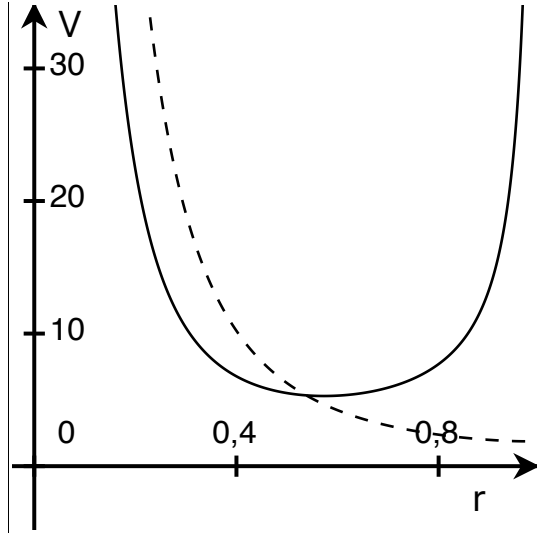}
\caption{Plots of potential (\ref{eq:ext-V-1-bis}) (solid line) and of potential (\ref{eq:SE-def-1}) (dashed line), both with $\kappa=L=Q=1$.}
\end{center}
\end{figure}
\par
%
%-------------------------------------------------------------------------------------------------------------------------------
%
\begin{figure}[h]
\begin{center}
\includegraphics{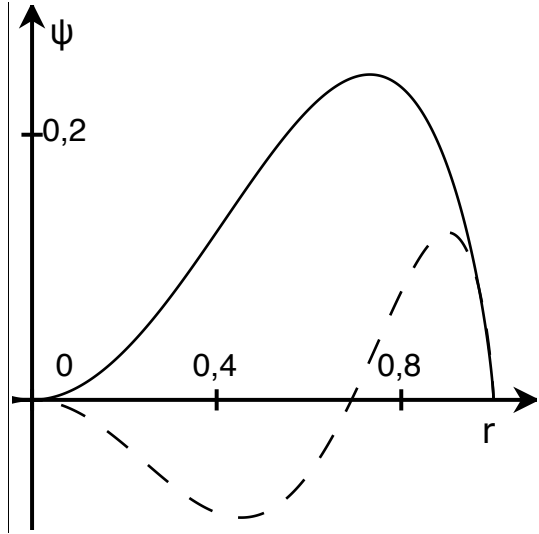}
\caption{Plots of ground state wavefunction $\psi_0(r)$ (solid line) and of first-excited state wavefunction $\psi_1(r)$ (dashed line) for potential (\ref{eq:ext-V-1-bis}) with $\kappa = L = Q = 1$. The corresponding eigenvalues are $E_0=15.9375$ and $E_1=35.9722$.}
\end{center}
\end{figure}
\par
%
%==============================================================
%
\subsection{Second extension}

{}For $m=2$, the potential (\ref{eq:ext-V}) reads
\begin{equation}
  V(r) = \frac{L(L+1)}{r^2} - \frac{Q}{r} f + \kappa B_1 \frac{r}{f} + \kappa B_2 \frac{1}{f^2} + \kappa B_3 
  \frac{r}{f^3} + \kappa B_4 \frac{1}{f^4}, \qquad B_4>0,
\end{equation}
and we assume a superpotential of the type
\begin{equation}
  W(r) = \frac{\xi}{r} f + \eta + \zeta \frac{r}{f} + \frac{\sigma}{f^2}, \qquad \xi \le 0, \qquad \sigma>0,
\end{equation}
depending on four parameters $\xi$, $\eta$, $\zeta$, and $\sigma$. From the Riccati equation for $W(r)$, we get the system of equations
\begin{equation}
\begin{split}
  &\xi(\xi+1) = L(L+1), \\
  &- \kappa \xi^2 + 2\xi\zeta + \eta^2 - \frac{\zeta^2}{\kappa} = - E_0, \\
  &2\xi(\eta+\sigma) = - Q, \\
  &2(\eta\zeta + \kappa\xi\sigma) = \kappa B_1, \\
  &2\eta\sigma + \frac{\zeta(\zeta-\kappa)}{\kappa} = \kappa B_2, \\
  &2\sigma(\zeta - \kappa) = \kappa B_3, \\
  &\sigma^2 = \kappa B_4.
\end{split}
\end{equation}
The latter leads to the values of the superpotential parameters
\begin{equation}
  \xi = -L-1, \quad \eta = \frac{Q}{2(L+1)} - \sqrt{\kappa B_4}, \quad \zeta = 
  \kappa \left(\frac{B_3}{2\sqrt{\kappa B_4}} + 1\right), \quad \sigma = \sqrt{\kappa B_4},
\end{equation}
the ground-state energy
\begin{equation}
  E_0 = \kappa \left(\frac{B_3}{2 \sqrt{\kappa B_4}} + L +2\right)^2 - \left(\frac{Q}{2(L+1)} - 
  \sqrt{\kappa B_4}\right)^2,  \label{eq:gs-E}
\end{equation}
and two constraints
\begin{equation}
\begin{split}
  & B_1 = 2\left(\frac{Q}{2(L+1)} - \sqrt{\kappa B_4}\right)\left(\frac{B_3}{2\sqrt{\kappa B_4}} + 1\right)
        - 2(L+1) \sqrt{\kappa B_4}, \\
  & B_2 = \frac{1}{\kappa} \left[2\left(\frac{Q}{2(L+1)} - \sqrt{\kappa B_4}\right) \sqrt{\kappa B_4} + \kappa
        \left(\frac{B_3}{2\sqrt{\kappa B_4}} + 1\right) \frac{B_3}{2\sqrt{\kappa B_4}}\right], 
\end{split}. \label{eq:constraint-1-bis}
\end{equation}
connecting the potential parameters. To $E_0$, given in (\ref{eq:gs-E}), corresponds the (normalizable) ground-state wavefunction
\begin{equation}
  \psi_0(r) \propto r^{-\xi} \exp\left(- \frac{\eta}{\sqrt{\kappa}} \arcsin(\sqrt{\kappa r}) - \sigma \frac{r}{f}
  \right) f^{\frac{\zeta}{\kappa}-\frac{1}{2}}.
\end{equation}
\par
%
%--------------------------------------------------------------------------------------------
%
The partner potential can be written as
\begin{equation}
  V_2(r) = \frac{L'(L'+1)}{r^2} - \frac{Q'}{r}f + \kappa B'_1 \frac{r}{f} + \kappa B'_2 \frac{1}{f^2} + \kappa
  B'_3 \frac{r}{f^3} + \kappa B'_4 \frac{1}{f^4} + R,
\end{equation}
where
\begin{equation}
\begin{split}
  &L'=L+1, \qquad Q'=Q, \qquad B'_1=B_1, \qquad B'_2= B_2 + \frac{B_3}{\sqrt{\kappa B_4}} + 2, \\
  &B'_3 = B_3 + 4 \sqrt{\kappa B_4}, \qquad B'_4 = B_4, \qquad R = - E_0.
\end{split}
\end{equation}
Hence, $V_1(r)$ is CDSI with the two constraints  (\ref{eq:constraint-1-bis}).\par
%
%--------------------------------------------------------------------------------------------------------------
%
On taking now a new superpotential in the form
\begin{equation}
  W'(r) = \frac{\xi'}{r}f + \eta' + \zeta' \frac{r}{f} + \frac{\sigma'}{f^2}, \qquad \xi' \le 0, \qquad \sigma'>0,
\end{equation}
and proceeding as in the previous subsection, we get the relations
\begin{equation}
  \xi' = -L-2, \quad \eta' = \frac{Q}{2(L+2)} - \sqrt{\kappa B_4}, \quad \zeta' = \kappa \left(
  \frac{B_3}{2\sqrt{\kappa B_4}} + 3\right), \quad \sigma' = \sqrt{\kappa B_4},
\end{equation}
and
\begin{equation}
  E'_0 = \kappa \left(\frac{B_3}{2\sqrt{\kappa B_4}} + L +5\right)^2 - \left(\frac{Q}{2(L+2)} - \sqrt{\kappa
  B_4}\right)^2,
\end{equation}
together with two new constraints
\begin{equation}
\begin{split}
  &B_1 = 2 \left(\frac{Q}{2(L+2)} - \sqrt{\kappa B_4}\right)\left(\frac{B_3}{2\sqrt{\kappa B_4}} + 3\right)
       - 2(L+2) \sqrt{\kappa B_4}, \\
  &B_2 = \frac{1}{\kappa} \left\{2\left(\frac{Q}{2(L+2)} - \sqrt{\kappa B_4}\right) \sqrt{\kappa B_4}
       + \kappa \left[\left(\frac{B_3}{2\sqrt{\kappa B_4}}\right)^2 + 3 \frac{B_3}{2\sqrt{\kappa B_4}} + 4\right]
       \right\},
\end{split} \label{eq:constraint-2-bis}
\end{equation}
\par
%
%--------------------------------------------------------------------------------------------------------------
%
The two sets of constraints (\ref{eq:constraint-1-bis}) and (\ref{eq:constraint-2-bis}) are compatible for some specific values of $B_3$ and $B_4$, which, in terms of the parameters $L$ and $\Q$, are given by
\begin{equation}
  B_3 = \frac{2\kappa (2L+3)\Q[(2L+1)\Q^2-6\kappa]}{(\Q^2+3\kappa)^2}, \qquad B_4 = \frac{\kappa
  (2L+3)^2\Q^2}{(\Q^2+3\kappa)^2}.
\end{equation}
The resulting values of $B_1$ and $B_2$ are then obtained as
\begin{equation}
\begin{split}
  &B_1 = 2\Q \frac{2(L+1)(L+2)\Q^4 - 3\kappa\Q^2 - 6\kappa^2(L^2+3L+3)}{(\Q^2+3\kappa)^2}, \\
  &B_2 = \frac{2(4L^2+10L+7)\Q^4 + \kappa(4L^2+3)\Q^2 + 18\kappa^2}{(\Q^2+3\kappa)^2}.
\end{split}
\end{equation}
The corresponding two superpotentials can be written as
\begin{equation}
\begin{split}
  W(r) &= - \frac{L+1}{r}f + \frac{\Q[(L+2)\Q^2+\kappa(L+3)]}{\Q^2+3\kappa} + \frac{\kappa[2(L+1)\Q^2
       -3\kappa]}{\Q^2+3\kappa} \frac{r}{f} \\ 
  &\quad{}+ \frac{\kappa(2L+3)\Q}{\Q^2+3\kappa} \frac{1}{f^2}, \\
  W'(r) &= - \frac{L+2}{r}f + \frac{\Q[(L+1)\Q^2+\kappa L]}{\Q^2+3\kappa} + \frac{\kappa[2(L+2)\Q^2
       +3\kappa]}{\Q^2+3\kappa} \frac{r}{f} \\
  &\quad{} + \frac{\kappa(2L+3)\Q}{\Q^2+3\kappa} \frac{1}{f^2}.
\end{split}. \label{eq:ext-W-bis}
\end{equation}
\par
%
%-------------------------------------------------------------------------------------------------------------------------
%
The potential $V(r)$ has a ground state and a first-excited state, whose energies are given by
\begin{equation}
\begin{split}
  &E_0 = 9\kappa \left(\frac{(L+1)\Q^2+\kappa L}{\Q^2+3\kappa}\right)^2 - \Q^2 \left(\frac{(L+2)\Q^2
      +\kappa(L+3)}{\Q^2+3\kappa}\right)^2, \\
  &E_1 = 9\kappa \left(\frac{(L+2)\Q^2 + \kappa(L+3)}{\Q^2+3\kappa}\right)^2 - \Q^2 \left(\frac{(L+1)\Q^2
      +\kappa L}{\Q^2+3\kappa}\right)^2,
\end{split} \label{eq:E-bis}
\end{equation}
respectively, while their corresponding normalizable wavefunctions are given by
\begin{align}
  \psi_0(r) &\propto r^{L+1} f^{\frac{(4L+3)\Q^2-9\kappa}{2(\Q^2+3\kappa)}} \exp\biggl(- \frac{\Q}
       {\sqrt{\kappa}}
       \frac{(L+2)\Q^2+ \kappa(L+3)}{\Q^2+3\kappa} \arcsin(\sqrt{\kappa} r) \nonumber \\
  & \quad{}- \frac{\kappa(2L+3)\Q}{\Q^2+3\kappa} \frac{r}{f}\biggr)
\end{align}
and
\begin{align}
  \psi_1(r) &\propto r^{L+1} f^{\frac{(4L+3)\Q^2-9\kappa}{2(\Q^2+3\kappa)}} \exp\biggl(- \frac{\Q}
       {\sqrt{\kappa}}
       \frac{(L+1)\Q^2+ \kappa L}{\Q^2+3\kappa} \arcsin(\sqrt{\kappa} r) \nonumber \\
  & \quad{}- \frac{\kappa(2L+3)\Q}{\Q^2+3\kappa} \frac{r}{f}\biggr) \left[\left(-1 + 3 \frac{\Q^2+\kappa}
       {\Q^2+3\kappa} \kappa r^2\right)f + \Q r\left(1 - \frac{\Q^2+\kappa}{\Q^2+3\kappa} \kappa r^2\right)
       \right].
\end{align}
\par
%
%+++++++++++++++++++++++++++++++++++++++++++++++++++++++++++++++++
%
\subsection{Higher extensions}

{}For higher $m$ values, the calculations are similar. On assuming a superpotential depending on $m+2$ parameters, the Riccati equation for $V_1$ provides a system of $2m+3$ equations, from which one gets the values of the superpotential parameters and of the ground-state energy, plus a set of $m$ constraints connecting the potential parameters. On repeating the procedure for the partner potential, one ends up with two sets of $m$ constraints, whose compatibility leads to the final parameter values. As an example, the results obtained for $m=3$ are listed in Appendix A.\par
%
%-----------------------------------------------------------------------------------------------
%
Since the complexity of the CDSI method considerably increases with the $m$ value, it is useful to explore the possibilities offered by the alternative method available for building potentials with two known eigenstates. This is the purpose of Section~4.\par
%
%==============================================================
%
\section{Generating function method}

\setcounter{equation}{0}

Let us start from two pairs of DSUSY partner Hamiltonians $(\hat{H}_1, \hat{H}_2)$ and $(\hat{H}'_1, \hat{H}'_2)$, defined as in (\ref{eq:partner-H}) and (\ref{eq:partner-H-bis}) in terms of two superpotentials $W(r)$ and $W'(r)$. The latter are related by the equation
\begin{equation}
  W^2(r) + f(r) \frac{dW(r)}{dr} = W^{\prime2}(r) - f(r) \frac{dW'(r)}{dr} + E_1 - E_0,  \label{eq:W-W'}
\end{equation}
which directly follows from (\ref{eq:V'-V}), where we set $E_1=E'_0$. From $W(r)$ and $W'(r)$, we may define the two functions
\begin{equation}
  W_+(r) = W'(r) + W(r), \qquad W_-(r) = W'(r) - W(r),  \label{eq:W_+-W_-}
\end{equation}
which allow us to rewrite Eq.~(\ref{eq:W-W'}) as
\begin{equation}
  f(r) \frac{dW_+(r)}{dr} = W_+(r) W_-(r) + E_1 - E_0.  \label{eq:gen-funct}
\end{equation}
Hence, $W_-(r)$ can be expressed in terms of $W_+(r)$ and the energy difference $E_1 - E_0$ as
\begin{equation}
  W_-(r) = \frac{f(r) dW_+(r)/dr + E_0 - E_1}{W_+(r)}.  \label{eq:W_-}
\end{equation}
\par
%
%----------------------------------------------------------------------------------------------------------
%
The generating function method starts from two functions $W_+(r)$ and $W_-(r)$ that are compatible, i.e., such that Eq.~(\ref{eq:gen-funct}) is satisfied for some positive constant $E_1 - E_0$. The two superpotentials are then determined from Eq.~(\ref{eq:W_+-W_-}). The starting potential $V_1(r)$ [as well as its partner $V_2(r)$] follows from Eq.~(\ref{eq:partner-H}) and its ground-state wavefunction from Eq.~(\ref{eq:gs}). Its first-excited state wavefunction, given in Eq.~(\ref{eq:es}) with $\hat{A}^+$ and $\psi'_0(r)$ defined in (\ref{eq:A}) and (\ref{eq:gs-bis}), respectively, can be rewritten as
\begin{equation}
  \psi_1(r) \propto W_+(r) f^{-1/2} \exp\left(- \int^r \frac{W'(r')}{f(r')} dr'\right),  \label{eq:es-bis}
\end{equation}
where use is made of Eq.~(\ref{eq:W_+-W_-}). It then remains to check that the functions $\psi_0(r)$ and $\psi_1(r)$ are normalizable functions on the definition interval of $r$.\par
%
%-----------------------------------------------------------------------------------------------------
%
The difficulty of the generating function method is to guess an appropriate function $W_+(r)$ [and its accompanying function $W_-(r)$], such that Eq.~(\ref{eq:gen-funct}) is satisfied for some positive constant $E_1-E_0$ and that the potential and wavefunctions are well behaved.\par
%
%++++++++++++++++++++++++++++++++++++++++++++++++++++++++
%
\subsection{Application to extensions of the Kepler-Coulomb potential on the sphere}

{}From the superpotentials $W(r)$ and $W'(r)$ obtained in Eqs.~(\ref{eq:ext-W}), (\ref{eq:ext-W-bis}), and (\ref{eq:ext-W-ter}) for the first three members of the extended KC potential family (\ref{eq:ext-V}), we obtain for the functions $W_+(r)$ and $W_-(r)$ the results
\begin{equation}
\begin{split}
  W_+(r) &= (2L+3) \left(- \frac{f}{r} + \Q + \kappa \frac{r}{f}\right), \\
  W_-(r) &= - \frac{f}{r} - \Q  + \kappa \frac{r}{f},
\end{split}
\end{equation}
if $m=1$,
\begin{equation}
\begin{split}
  W_+(r) &= (2L+3) \left(- \frac{f}{r} + \Q \frac{\Q^2+\kappa}{\Q^2+3\kappa} + \frac{2\kappa\Q^2}
      {\Q^2+3\kappa} \frac{r}{f} + \frac{2\kappa\Q}{\Q^2+3\kappa} \frac{1}{f^2}\right), \\
  W_-(r) &= - \frac{f}{r} - \Q + 2\kappa \frac{r}{f},
\end{split}
\end{equation}
if $m=2$, and
\begin{equation}
\begin{split}
  W_+(r) &= (2L+3) \biggl(- \frac{f}{r} + \Q \frac{\Q^2+4\kappa}{\Q^2+10\kappa} + 3\kappa 
       \frac{\Q^2+2\kappa}{\Q^2+10\kappa} \frac{r}{f} + \frac{6\kappa\Q}{\Q^2+10\kappa} \frac{1}{f^2} \\
  & \quad{}+ \frac{6\kappa^2}{\Q^2+10\kappa} \frac{r}{f^3}\biggr), \\
  W_-(r) &= - \frac{f}{r} - \Q + 3\kappa \frac{r}{f},
\end{split}
\end{equation}
if $m=3$. These functions satisfy Eq.~(\ref{eq:gen-funct}) with the constant $E_1 - E_0$ derived from Eqs.~(\ref{eq:E}), (\ref{eq:E-bis}), and (\ref{eq:E-ter}), and given by
\begin{equation}
\begin{split}
  E_1 - E_0 &= (2L+3) (\Q^2+4\kappa), \\
  E_1 - E_0 &= (2L+3) \frac{(\Q^2+\kappa)(\Q^2+9\kappa)}{\Q^2+3\kappa}, \\
  E_1 - E_0 &= (2L+3) \frac{(\Q^2+4\kappa)(\Q^2+16\kappa)}{\Q^2+10\kappa},
\end{split}
\end{equation}
respectively.\par
%
%--------------------------------------------------------------------------------------------------------
%
As a generalization for any $m \in \N^+$, we propose to consider
\begin{equation}
\begin{split}
  &W_+(r) = (2L+3) \left(- \frac{f}{r} + \Q a_0 + \kappa \sum_{k=0}^{[(m-1)/2]} a_{2k+1} \frac{r}{f^{2k+1}}
        + \sum_{k=1}^{[m/2]} a_{2k} \frac{1}{f_{2k}}\right), \\
  &W_-(r) = - \frac{f}{r} - \Q + m\kappa \frac{r}{f}, \\
  &E_1 - E_0 = (2L+3) a_0 [\Q^2 + (m+1)^2\kappa],
\end{split}. \label{eq:gen-m}
\end{equation}
where $a_0$, $a_1$, \ldots, $a_m$ are $m+1$ constants to be determined in such a way that Eq.~(\ref{eq:gen-funct}) is satified. These generalized formulas agree with the results previously obtained for $m=1$, 2, and 3 if the parameters $a_0$, $a_1$, \ldots, $a_m$ are chosen as follows:
\begin{equation}
\begin{split}
  a_0 &= 1, \quad a_1 = 1, \quad \text{if $m=1$}, \\
  a_0 &= \frac{\Q^2+\kappa}{\Q^2+3\kappa}, \quad a_1 = \frac{2\Q^2}{\Q^2+3\kappa}, \quad a_2 =
      \frac{2\kappa\Q}{\Q^2+3\kappa}, \quad \text{if $m=2$}, \\
  a_0 &= \frac{\Q^2+4\kappa}{\Q^2+10\kappa}, \quad a_1 = \frac{3(\Q^2+2\kappa)}{\Q^2+10\kappa}, \quad
      a_2 = \frac{6\kappa\Q}{\Q^2+10\kappa}, \quad a_3 = \frac{6\kappa}{\Q^2+10\kappa}, \\
  &\qquad{} \text{if $m=3$}.
\end{split}. \label{eq:a}
\end{equation}
\par
%
%++++++++++++++++++++++++++++++++++++++++++++++++++++++++++++++
%
\subsection{\boldmath Results for higher $m$ values}

To find the values of the $a_i$ coefficients in (\ref{eq:gen-m}), let us insert the explicit expressions for $W_+(x)$, $W_-(x)$, and $E_1-E_0$ in Eq.~(\ref{eq:gen-funct}) and use the following identities:
\begin{equation}
\begin{split}
  &\frac{df}{dr} = - \kappa \frac{r}{f}, \\
  &\kappa r^2 = 1 - f^2, \\
  &\frac{1}{r f^{2k}} = \frac{1}{r} + \kappa \left(\frac{r}{f^2} + \frac{r}{f^4} + \cdots + \frac{r}{f^{2k}}\right),
       \quad k=1, 2, \ldots, \\
  &\frac{1}{r f^{2k+1}} = \frac{f}{r} + \kappa \left(\frac{r}{f} + \frac{r}{f^3} + \cdots + \frac{r}{f^{2k+1}}
       \right), \quad k=0, 1, 2, \ldots.
\end{split}
\end{equation}
On both sides of (\ref{eq:gen-funct}), we then obtain a linear combination of 1, $r/f$, $r/f^{2k+1}$ ($1\le k \le [(m-1)/2]$), $1/f^{2k}$ ($1 \le k \le [m/2]$), and $f/r$. On equating their coefficients, we get the relations
\begin{align}
  &(m+1) a_0 - a_1 = 1, \label{eq:cond-1}\\
  &m\Q a_0 - \Q a_1 - \sum_{k=1}^{[m/2]} a_{2k} = 0, \label{eq:cond-2} \\
  &(m-2k) a_{2k} - \Q a_{2k+1} - \sum_{l=k+1}^{[m/2]} a_{2l} = 0, \quad k=1, 2, \ldots, \left[\frac{m-1}{2}
       \right], \label{eq:cond-3} \\
  &\kappa(m+1-2k) a_{2k-1} - \Q a_{2k} - \kappa(m+1-2k) a_{2k+1} = 0, \quad k=1, 2, \ldots, \left[\frac{m}{2}
       \right] -  1,  \label{eq:cond-4} \\
  &\frac{1}{2}\bigl(1- (-1)^m\bigr) (2\kappa a_{m-2} - \Q a_{m-1} - 2\kappa a_m) + \frac{1}{2}\bigl(1 +
       (-1)^m\bigr) (\kappa a_{m-1} - \Q a_m) = 0,  \label{eq:cond-5} \\
  &\Q - \Q a_0 - \sum_{k=1}^{[m/2]} a_{2k} = 0,  \label{eq:cond-6}  
\end{align}
which form a system of $m+1$ linear equations in the $m+1$ unknowns $a_0$, $a_1$, \ldots, $a_m$ because the last equation (\ref{eq:cond-6}) turns out to be a combination of Eqs.~(\ref{eq:cond-1}) and (\ref{eq:cond-2}). The solutions of this system of equations for some values of $m$ higher than three are given in Appendix~B.\par
%
%--------------------------------------------------------------------------------------------------------------------
%
{}From the results  obtained for $m \le 7$ in (\ref{eq:a}) and in Appendix~B, we feel justified in formulating the following conjecture.

\noindent
{\bf Conjecture} For any $m>1$, the solution of the system of equations (\ref{eq:cond-1})--(\ref{eq:cond-5}) can be written as
\begin{align}
  a_0 &= \frac{(\Q^2+4\kappa)(\Q^2+16\kappa) \ldots (\Q^2+4\mu^2\kappa)}{\Q^{2\mu} + b_1\kappa
       \Q^{2\mu-2} + b_2\kappa^2\Q^{2\mu-4} + \cdots + b_{\mu}\kappa^{\mu}}, \\
  a_{2k} &= \frac{(2\mu+1)!}{(2\mu-2k+1)!} \frac{\kappa^k \Q (\Q^2+4\kappa)(\Q^2+16\kappa) \ldots
       (\Q^2+4(\mu-k)^2\kappa)}{\Q^{2\mu} + b_1\kappa\Q^{2\mu-2} + b_2\kappa^2\Q^{2\mu-4} + \cdots 
       + b_{\mu}\kappa^{\mu}}, \nonumber \\
  &\quad{} k=1, 2, \ldots, \mu, \\
  a_{2k+1} &= \frac{(2\mu+1)!}{(2\mu-2k)!} \frac{\kappa^k (\Q^{2\mu-2k} + c^{(k)}_1\kappa\Q^{2\mu-2k-2}
       + c^{(k)}_2\kappa^2\Q^{2\mu-2k-4} + \cdots + c^{(k)}_{\mu-k}\kappa^{\mu-k})}{\Q^{2\mu} + 
       b_1\kappa\Q^{2\mu-2} + b_2\kappa^2\Q^{2\mu-4} + \cdots + b_{\mu}\kappa^{\mu}}, \nonumber \\
  &\quad{} k=0, 1, \ldots, \mu, 
\end{align}
if $m=2\mu+1$, and 
\begin{align}
  a_0 &= \frac{(\Q^2+\kappa)(\Q^2+9\kappa) \ldots (\Q^2+(2\mu-1^2\kappa)}{\Q^{2\mu} + b_1\kappa
       \Q^{2\mu-2} + b_2\kappa^2\Q^{2\mu-4} + \cdots + b_{\mu}\kappa^{\mu}}, \\
  a_{2k} &= \frac{(2\mu)!}{(2\mu-2k)!} \frac{\kappa^k\Q(\Q^2+\kappa)(\Q^2+9\kappa) \ldots (\Q^2
       +(2\mu-2k-1)^2\kappa)}{\Q^{2\mu} + b_1\kappa
       \Q^{2\mu-2} + b_2\kappa^2\Q^{2\mu-4} + \cdots + b_{\mu}\kappa^{\mu}}, \nonumber \\
  &\quad{} k=1, 2, \ldots, \mu, \\
  a_{2k+1} &= \frac{(2\mu)!}{(2\mu-2k-1)!} \nonumber \\
  &\quad{} \times \frac{\kappa^k\Q^2(\Q^{2\mu-2k-2}+c^{(k)}_1\kappa
       \Q^{2\mu-2k-4} + c^{(k)}_2\kappa^2\Q^{2\mu-2k-6} + \cdots + c^{(k)}_{\mu-k-1}\kappa^{\mu-k-1}}
       {\Q^{2\mu} + b_1\kappa\Q^{2\mu-2} + b_2\kappa^2\Q^{2\mu-4} + \cdots + b_{\mu}\kappa^{\mu}}, 
       \nonumber \\
  &\quad{} k=0,1, \ldots, \mu-1,
\end{align}
if $m=2\mu$, where $b_1$, $b_2$, \ldots, $b_{\mu}$, and $c^{(k)}_1$, $c^{(k)}_2$, \ldots, $c^{(k)}_{\mu-k}$ or $c^{(k)}_{\mu-k-1}$ are some positive constants.\par
%
%--------------------------------------------------------------------------------------------------------
%
{}From (\ref{eq:W_+-W_-}) and (\ref{eq:gen-m}), the superpotential $W(r)$ assumes the form
\begin{align}
  W(r) &= \frac{1}{2} \biggl\{- (2L+2) \frac{f}{r} + \Q[(2L+3)a_0+1] + \kappa [(2L+3)a_1-m] \frac{r}{f}
      \nonumber \\
  &\quad{} + \kappa (2L+3) \sum_{k=1}^{[(m-1)/2]} a_{2k+1} \frac{r}{f^{2k+1}} + (2L+3) \sum_{k=1}
      ^{[m/2]} a_{2k} \frac{1}{f^{2k}}\biggr\}
\end{align}
and from the equation $W^2 - f dW/dr = V - E_0$, resulting from (\ref{eq:rescaled-V}) and (\ref{eq:partner-H}), one can easily check that $V(r)$ is an extended KC potential, in other words it contains a term $L(L+1)/r^2$, as well as a term
\begin{align}
  & \frac{1}{4} \biggl\{- 2(2L+2) \Q [(2L+3) a_0 + 1] - 2(2L+2)(2L+3) \sum_{k=1}^{[m/2]} a_{2k} 
        \biggr\}\frac{f}{r} \nonumber \\
  &\quad = -2(L+1)(L+2) \Q \frac{f}{r} = - Q \frac{f}{r},
\end{align}
where use has been made of Eq.~(\ref{eq:cond-6}). The other parameters $B_1$, $B_2$, \ldots, $B_{2m}$ of Eq.~(\ref{eq:ext-V}) can in principle be calculated from the same equation, but their explicit expressions in terms of the parameters $a_i$ being very complicated, we will not give them here. In contrast, the constant term in $W^2 - fdW/dr$ easily provides the value of $-E_0$ in terms of $a_0$ and $a_1$,
\begin{align}
  -E_0 &= \frac{1}{4} \biggl\{- (2L+2)^2 \kappa + \Q^2 [(2L+3)a_0+1]^2 - [(2L+3)a_1-m^2] \kappa
       \nonumber \\
  & \quad{} - 2(2L+2)[(2L+3)a_1-m]\kappa\biggr\}.
\end{align}
On expressing $a_1$ in terms of $a_0$ through Eq.~(\ref{eq:cond-1}), we get
\begin{equation}
  E_0 = \frac{1}{4} \left\{- \Q^2 [(2L+3)a_0+1]^2 + (m+1)^2 \kappa [(2L+3)a_0-1]^2\right\}.
\end{equation}
Furthermore, on combining this result with Eq.~(\ref{eq:gen-m}), we obtain for the first-excited state energy
\begin{equation}
  E_1 = \frac{1}{4} \left\{- \Q^2 [(2L+3)a_0-1]^2 + (m+1)^2 \kappa [(2L+3)a_0+1]^2\right\}.
\end{equation}
We conclude that the first two eigenvalues of $V(r)$ are expressed in terms of the single parameter $a_0$.\par
%
%--------------------------------------------------------------------------------------------------------------------
%
The corresponding wavefunctions, determined from Eqs.~(\ref{eq:gs}) and (\ref{eq:es-bis}), are given by
\begin{align}
  \psi_0(r) &\propto r^{L+1} f^{\frac{1}{2}[(2L+3)a_1-m-1]} \nonumber \\
  & \quad{} \times \exp\biggl\{- \frac{\Q}{2\sqrt{\kappa}} [(2L+3)a_0+1]\arcsin(\sqrt{\kappa} r)
       - \frac{1}{4}(2L+3) \sum_{k=1}^{[(m-1)/2]} \frac{a_{2k+1}}{kf^{2k}} \nonumber \\
  & \quad {}- \frac{1}{2}(2L+3) \sum_{k=1}^{[m/2]} \frac{(2k-3)!!}{(2k-2)!!} \frac{r}{f^{2k-1}}
       \sum_{l=k}^{[m/2]} \frac{(2l-2)!!}{(2l-1)!!} a_{2l}\biggr\} 
\end{align}
and
\begin{align}
  \psi_1(r) &\propto W_+(r) r^{L+2} f^{\frac{1}{2}[(2L+3)a_1+m-1]} \nonumber \\
  & \quad{} \times \exp\biggl\{- \frac{\Q}{2\sqrt{\kappa}} [(2L+3)a_0-1]\arcsin(\sqrt{\kappa} r)
       - \frac{1}{4}(2L+3) \sum_{k=1}^{[(m-1)/2]} \frac{a_{2k+1}}{kf^{2k}} \nonumber \\
  & \quad {}- \frac{1}{2}(2L+3) \sum_{k=1}^{[m/2]} \frac{(2k-3)!!}{(2k-2)!!} \frac{r}{f^{2k-1}}
       \sum_{l=k}^{[m/2]} \frac{(2l-2)!!}{(2l-1)!!} a_{2l}\biggr\}, 
\end{align}
respectively. They are normalizable on the interval $(0, 1/\sqrt{\kappa})$ provided $a_m > 0$, which is the case for all $m$ values for which explicit calculations have been carried out.\par
%
%===============================================================
% 
\section{Conclusion}

In the present paper, we have shown that the combination  of methods previously used to build families of extended oscillator potentials in a curved space with known ground and first-excited states can be applied to extended KC potentials on the sphere containing $2m$ additional terms, where $m$ may take any value in $\N^+$.\par
%
%--------------------------------------------------------------------------------------------------------------
%
We have first extended the DSI symmetry, known to be valid for the KC potential alone, by completing it with $m$ constraints relating the parameters of the extended potential, which therefore turns out to be CDSI. The second step in the construction of an extended potential hierarchy has then provided us with a second set of $m$ constraints among the extended potential parameters. Compatibility conditions between the two sets of $m$ constraints have given rise to QES extensions of the KC potential with two known eigenstates. In this way, some explicit results have been obtained for the first three members of the extended KC potential family.\par
%
%---------------------------------------------------------------------------------------------------------------
%
To generalize those outcomes to other members of the extended KC potential family, we have then turned ourselves to the generating function method, wherein the first two DSUSY superpotentials (and hence the potentials and their first two eigenstates) are expressed in terms of a function $W_+(r)$ [and its accompanying function $W_-(r)$]. From the results obtained for $W_{\pm}(r)$ for the first three members of the family, we have proposed some general formulas for $W_{\pm}(r)$, depending on $m+1$ constants $a_0$, $a_1$, \ldots, $a_m$, and we have shown that such constants must satisfy a system of $m+1$ linear equations. On solving such a system, we have obtained some explicit results for $m=4$, 5, 6, and 7. We have then formulated a conjecture giving the general structure of the $a_i$ constants in terms of the parameters of the problem. All the aspects of this conjecture and the consequences drawn from it for the first two eigenstates of $V(r)$ have been confirmed for $m\le 7$.\par
%
%-------------------------------------------------------------------------------------------------------
%
The problem that remains open is to prove our conjecture by finding the explicit values of the constants appearing there. Solving it would be an interesting topic for future investigation. Applying the methods presented here to extensions of the KC potential in a hyperbolic space would also be an open question for future work.\par
%
%-----------------------------------------------------------------------------------------------------------
%
Taking into account the large number of physical problems wherein the KC potential (either in curved space or in a PDM background) provides a first crude approximation and some additional terms are needed to reproduce experimental data, it is obvious that the exact results presented here for potentials including some additional terms may find some interesting applications to practical problems. The search for such applications would be another interesting topic for future investigation.\par
%
%========================================================================
%
\section*{Acknowledgments}

The author was supported  by the Fonds de la Recherche Scientifique - FNRS under Grant Number 4.45.10.08.\par
%
%================================================================
%
\section*{Author declarations}

\subsection*{Conflicts of interest}

The author declares no conflict of interest.\par
%
%=================================================================
%
\section*{Data availability}

All data supporting the findings of this study are included in the article.\par
%
%=================================================================
%
\section*{Appendix A: Results for the third extension of the Kepler-Coulomb potential}

\renewcommand{\theequation}{A.\arabic{equation}}
\setcounter{equation}{0}

The third extension of the KC potential reads
\begin{equation}
  V(r) = \frac{L(L+1)}{r^2} - \frac{Q}{r} f + \kappa B_1 \frac{r}{f} + \kappa B_2 \frac{1}{f^2} + \kappa B_3
  \frac{r}{f^3} + \kappa B_4 \frac{1}{f^4} + \kappa B_5 \frac{r}{f^5} + \kappa B_6 \frac{1}{f^6}.
  \label{eq:V-ter}
\end{equation}
The CDSI method leads to the following values of the coefficients:
\begin{equation}
\begin{split}
  B_1 &= 6\Q \frac{(L+1)(L+2)\Q^4 + 4\kappa(L^2+3L+1)\Q^2 - 4\kappa^2(3L^2+9L+13)}
      {(\Q^2+10\kappa)^2}, \\
  B_2 &= 3 \frac{(7L^2+19L+14)\Q^4 + 4\kappa(3L^2+5L+7)\Q^2 - 4\kappa^2(13L^2+29L-17)}
      {(\Q^2+10\kappa)^2}, \\
  B_3 &= \frac{6\kappa(2L+3)\Q [4(L+1)\Q^2 + 5\kappa(2L-1)]}{(\Q^2+10\kappa)^2},  \\
  B_4 &= \frac{9\kappa(2L+3) [4(L+1)\Q^2 + \kappa(2L-17)]}{(\Q^2+10\kappa)^2}, \\
  B_5 &= \frac{18\kappa^2(2L+3)^2\Q}{(\Q^2+10\kappa)^2}, \\
  B_6 &= \frac{9\kappa^2(2L+3)^2}{(\Q^2+10\kappa)^2}.
\end{split}. \label{eq:B-ter}
\end{equation}
%
%-------------------------------------------------------------------------------------------------------------
%
The corresponding two superpotentials are
\begin{equation}
\begin{split}
  W(r) &= - \frac{L+1}{r}f + \frac{\Q[(L+2)\Q^2+ \kappa(4L+11)]}{\Q^2+10\kappa} + \frac{3\kappa
       [(L+1)\Q^2 +2\kappa(L-1)]}{\Q^2+10\kappa} \frac{r}{f} \\
  &\quad + \frac{3\kappa(2L+3)\Q}{\Q^2+10\kappa} \frac{1}{f^2} + \frac{3\kappa^2(2L+3)}{\Q^2+10\kappa}
       \frac{r}{f^3},  \\ 
  W'(r) &= - \frac{L+2}{r}f + \frac{\Q[(L+1)\Q^2+ \kappa(4L+1)]}{\Q^2+10\kappa} + \frac{3\kappa
       [(L+2)\Q^2 +2\kappa(L+4)]}{\Q^2+10\kappa} \frac{r}{f} \\
  &\quad + \frac{3\kappa(2L+3)\Q}{\Q^2+10\kappa} \frac{1}{f^2} + \frac{3\kappa^2(2L+3)}{\Q^2+10\kappa}
       \frac{r}{f^3}.
\end{split}. \label{eq:ext-W-ter}
\end{equation}
\par
%
%--------------------------------------------------------------------------------------------------------------------------
%
The first two bound states of the potential defined in (\ref{eq:V-ter}) and (\ref{eq:B-ter}) are characterized by the energies
\begin{equation}
\begin{split}
  E_0 &= 16\kappa \left(\frac{(L+1)\Q^2+\kappa(4L+1)}{\Q^2+10\kappa}\right)^2 - \Q^2
       \left(\frac{(L+2)\Q^2+\kappa(4L+11)}{\Q^2+10\kappa}\right)^2, \\
  E_1 &= 16\kappa \left(\frac{(L+2)\Q^2+\kappa(4L+11)}{\Q^2+10\kappa}\right)^2 - \Q^2
       \left(\frac{(L+1)\Q^2+\kappa(4L+1)}{\Q^2+10\kappa}\right)^2,
\end{split} \label{eq:E-ter}
\end{equation}
and their wavefunctions can be written as
\begin{equation}
\begin{split}
  \psi_0(r) &\propto r^{L+1} f^{\frac{(6L+5)\Q^2+2\kappa(6L-11)}{2(\Q^2+10\kappa)}}
       \exp{\biggl(-\frac{\Q}{\sqrt{\kappa}} \frac{(L+2)\Q^2+\kappa(4L+11)}{\Q^2+10\kappa}\biggr)} \\
  &\quad{}\times \exp{\biggl( - \frac{3\kappa(2L+3)\Q}{\Q^2+10\kappa} \frac{r}{f} - \frac{3\kappa(2L+3)}
       {2(\Q^2+10\kappa)}\frac{1}{f^2}\biggr)},  \\
  \psi_1(r) &\propto r^{L+1} f^{\frac{(6L+5)\Q^2+2\kappa(6L-11)}{2(\Q^2+10\kappa)}}
       \exp{\biggl(-\frac{\Q}{\sqrt{\kappa}} \frac{(L+1)\Q^2+\kappa(4L+1)}{\Q^2+10\kappa}\biggr)} \\
  &\quad{}\times \exp{\biggl( - \frac{3\kappa(2L+3)\Q}{\Q^2+10\kappa} \frac{r}{f} - \frac{3\kappa(2L+3)}
       {2(\Q^2+10\kappa)}\frac{1}{f^2}\biggr)} \\
  &\quad{}\times \biggl[\Q\biggl(1 - \frac{\Q^2+4\kappa}{\Q^2+10\kappa} \kappa r^2\biggr) rf - 1
       + \frac{5\Q^2+32\kappa}{\Q^2+10\kappa} \kappa r^2 - 4 \frac{\Q^2+4\kappa}{\Q^2+10\kappa}
       \kappa^2 r^4\biggr],
\end{split}
\end{equation}
respectively.\par
%
%================================================================
%
\section*{\boldmath Appendix B: Solution of the system of equations (\ref{eq:cond-1})--(\ref{eq:cond-5}) and eigenvalues $E_0$ and $E_1$ for $m=4$, 5, 6, 7}

\renewcommand{\theequation}{B.\arabic{equation}}
\setcounter{equation}{0}

The solution of the system of $m+1$ equations (\ref{eq:cond-1})--(\ref{eq:cond-5}) is given by
\begin{align}
  a_0&= \frac{(\Q^2+\kappa)(\Q^2+9\kappa)}{\Q^4 + 22\kappa\Q^2 + 45\kappa^2}, \quad
       a_1 = \frac{4\Q^2(\Q^2+7\kappa)}{\Q^4 + 22\kappa\Q^2 + 45\kappa^2}, \quad
       a_2 = \frac{12\kappa\Q(\Q^2+\kappa)}{\Q^4 + 22\kappa\Q^2 + 45\kappa^2}, \nonumber \\
  a_3 &= \frac{24\kappa\Q^2}{\Q^4 + 22\kappa\Q^2 + 45\kappa^2}, \quad
       a_4 = \frac{24\kappa^2\Q}{\Q^4 + 22\kappa\Q^2 + 45\kappa^2},
\end{align}
if $m=4$,
\begin{align}
  a_0 &= \frac{(\Q^2+4\kappa)(\Q^2+16\kappa)}{\Q^4+40\kappa\Q^2+264\kappa^2}, \quad
      a_1 = \frac{5(\Q^4+16\kappa\Q^2+24\kappa^2)}{\Q^4+40\kappa\Q^2+264\kappa^2}, \nonumber \\
  a_2 &= \frac{20\kappa\Q(\Q^2+4\kappa)}{\Q^4+40\kappa\Q^2+264\kappa^2}, \quad
      a_3 = \frac{60\kappa(\Q^2+2\kappa)}{\Q^4+40\kappa\Q^2+264\kappa^2}, \nonumber \\
  a_4 &=. \frac{120\kappa^2\Q}{\Q^4+40\kappa\Q^2+264\kappa^2}, \quad
      a_5 = \frac{120\kappa^2}{\Q^4+40\kappa\Q^2+264\kappa^2},
\end{align}
if $m=5$,
\begin{align}
  a_0 &= \frac{(\Q^2+\kappa)(\Q^2+9\kappa)(\Q^2+25\kappa)}{\Q^6+65\kappa\Q^4+919\kappa^2\Q^2
      +1575\kappa^3}, \nonumber \\
  a_1 &= \frac{6\Q^2(\Q^4+30\kappa\Q^2+149\kappa^2)} 
      {\Q^6+65\kappa\Q^4+919\kappa^2\Q^2+1575\kappa^3}, \nonumber \\
  a_2 &= \frac{30\kappa\Q(\Q^2+\kappa)(\Q^2+9\kappa)}{\Q^6+65\kappa\Q^4+919\kappa^2\Q^2
      +1575\kappa^3}, \nonumber \\
  a_3 &= \frac{120\kappa\Q^2(\Q^2+7\kappa)}{\Q^6+65\kappa\Q^4+919\kappa^2\Q^2
      +1575\kappa^3}, \nonumber \\
  a_4 &= \frac{360\kappa^2\Q(\Q^2+\kappa)}{\Q^6+65\kappa\Q^4+919\kappa^2\Q^2 +1575\kappa^3},
      \nonumber \\
  a_5 &= \frac{720\kappa^2\Q^2}{\Q^6+65\kappa\Q^4+919\kappa^2\Q^2 +1575\kappa^3},
      \nonumber \\
  a_6 &= \frac{720\kappa^3\Q}{\Q^6+65\kappa\Q^4+919\kappa^2\Q^2+1575\kappa^3},
\end{align}
if $m=6$, and
\begin{align}
  a_0 &= \frac{(\Q^2+4\kappa)(\Q^2+16\kappa)(\Q^2+36\kappa)}{\Q^6+98\kappa\Q^4+2464\kappa^2\Q^2
       +13392\kappa^3}, \nonumber \\
  a_1 &= \frac{7(\Q^6+50\kappa\Q^4+544\kappa^2\Q^2+720\kappa^3)}
       {\Q^6+98\kappa\Q^4+2464\kappa^2\Q^2+13392\kappa^3}, \nonumber \\
  a_2 &= \frac{42\kappa\Q(\Q^2+4\kappa)(\Q^2+16\kappa)}{\Q^6+98\kappa\Q^4+2464\kappa^2\Q^2
       +13392\kappa^3}, \nonumber \\
  a_3 &= \frac{210\kappa(\Q^4+16\kappa\Q^2+24\kappa^2)}{\Q^6+98\kappa\Q^4+2464\kappa^2\Q^2
       +13392\kappa^3}, \nonumber \\
  a_4 &= \frac{840\kappa^2\Q(\Q^2+4\kappa)}{\Q^6+98\kappa\Q^4+2464\kappa^2\Q^2
       +13392\kappa^3}, \nonumber \\
  a_5 &= \frac{2520\kappa^2(\Q^2+2\kappa)}{\Q^6+98\kappa\Q^4+2464\kappa^2\Q^2
       +13392\kappa^3}, \nonumber \\
  a_6 &= \frac{5040\kappa^3\Q}{\Q^6+98\kappa\Q^4+2464\kappa^2\Q^2+13392\kappa^3}, \nonumber \\
  a_7 &= \frac{5040\kappa^3}{\Q^6+98\kappa\Q^4+2464\kappa^2\Q^2+13392\kappa^3},    
\end{align}
if $m=7$.\par
%
%----------------------------------------------------------------------------------------------------------------------
%
The corresponding eigenvalues of $V(r)$ are obtained in the form
\begin{align}
  E_0 &= 25\kappa \biggl(\frac{(L+1)\Q^4+2\kappa(5L+2)\Q^2+9\kappa^2(L-1)}
       {\Q^4+22\kappa\Q^2+45\kappa^2}\biggr)^2 \nonumber \\
  &\quad{} -\Q^2\biggl(\frac{(L+2)\Q^4+2\kappa(5L+13)\Q^2+9\kappa^2(L+4)}
       {\Q^4+22\kappa\Q^2+45\kappa^2}\biggr)^2, \nonumber \\
  E_1 &= 25\kappa\biggl(\frac{(L+2)\Q^4+2\kappa(5L+13)\Q^2+9\kappa^2(L+4)}
       {\Q^4+22\kappa\Q^2+45\kappa^2}\biggr)^2 \nonumber\\
  &\quad{} - \Q^2\biggl(\frac{(L+1)\Q^4+2\kappa(5L+2)\Q^2+9\kappa^2(L-1)}
       {\Q^4+22\kappa\Q^2+45\kappa^2}\biggr)^2, 
\end{align}
if $m=4$,
\begin{align}
  E_0 &= 36\kappa \biggl(\frac{(L+1)\Q^4+10\kappa(2L+1)\Q^2+32\kappa^2(2L-1)}
      {\Q^4+40\kappa\Q^2+264\kappa^2}\biggr)^2 \nonumber \\
  &\quad - \Q^2\biggl(\frac{(L+2)\Q^4+10\kappa(2L+5)\Q^2+32\kappa^2(2L+7)}
      {\Q^4+40\kappa\Q^2+264\kappa^2}\biggr)^2, \nonumber \\
  E_1 &= 36\kappa\biggl(\frac{(L+2)\Q^4+10\kappa(2L+5)\Q^2+32\kappa^2(2L+7)}
      {\Q^4+40\kappa\Q^2+264\kappa^2}\biggr)^2 \nonumber \\
  &\quad{} - \Q^2 \biggl(\frac{(L+1)\Q^4+10\kappa(2L+1)\Q^2+32\kappa^2(2L-1)}
      {\Q^4+40\kappa\Q^2+264\kappa^2}\biggr)^2,  
\end{align}
if $m=5$,
\begin{align}
  E_0 &= 49\kappa \biggl(\frac{(L+1)\Q^6+5\kappa(7L+4)\Q^4+\kappa^2(259L-71)\Q^2+225\kappa^3(L-2)}
      {\Q^6+65\kappa\Q^4+912\kappa^2\Q^2+1575\kappa^3}\biggr)^2 \nonumber \\
  &\quad - \Q^2 \biggl(\frac{(L+2)\Q^6+5\kappa(7L+17)\Q^4+\kappa^2(259L+848)\Q^2+225\kappa^3(L+5)}
      {\Q^6+65\kappa\Q^4+912\kappa^2\Q^2+1575\kappa^3}\biggr)^2, \nonumber \\
  E_1 &= 49\kappa\biggl(\frac{(L+2)\Q^6+5\kappa(7L+17)\Q^4+
       \kappa^2(259L+848)\Q^2+225\kappa^3(L+5)}
      {\Q^6+65\kappa\Q^4+912\kappa^2\Q^2+1575\kappa^3}\biggr)^2 \nonumber \\
  &\quad{} - \Q^2 \biggl(\frac{(L+1)\Q^6+5\kappa(7L+4)\Q^4+\kappa^2(259L-71)\Q^2+225\kappa^3(L-2)}
      {\Q^6+65\kappa\Q^4+912\kappa^2\Q^2+1575\kappa^3}\biggr)^2,
\end{align}
if $m=6$, and
\begin{align}
  E_0 &= 64\kappa \biggl(\frac{(L+1)\Q^6+7\kappa(8L+5)\Q^4+56\kappa^2(14L-1)\Q^2+72\kappa^3(32L
      -45)}{\Q^6+98\kappa\Q^4+2464\kappa^2\Q^2+13392\kappa^3}\biggr)^2 \nonumber \\
  &\quad - \Q^2 \biggl(\frac{(L+2)\Q^6+7\kappa(8L+19)\Q^4+56\kappa^2(14L+43)\Q^2+72\kappa^3(32L
      +141)}{\Q^6+98\kappa\Q^4+2464\kappa^2\Q^2+13392\kappa^3}\biggr)^2, \nonumber \\
  E_1 &= 64\kappa\biggl(\frac{(L+2)\Q^6+7\kappa(8L+19)\Q^4+56\kappa^2(14L+43)\Q^2+72\kappa^3(32L
      +141)}{\Q^6+98\kappa\Q^4+2464\kappa^2\Q^2+13392\kappa^3}\biggr)^2 \nonumber \\
  &\quad - \Q^2 \biggl(\frac{(L+1)\Q^6+7\kappa(8L+5)\Q^4+56\kappa^2(14L-1)\Q^2+72\kappa^3(32L
      -45)}{\Q^6+98\kappa\Q^4+2464\kappa^2\Q^2+13392\kappa^3}\biggr)^2, 
\end{align}
if $m=7$.\par
%
%=============================================================
%

\end{document}